\documentclass[pra,aps,showpacs,twocolumn,unsortedaddress,superscriptaddress,showpacs]{revtex4}
\usepackage{graphics,bm}
\usepackage{amssymb,amsmath}
\usepackage{epsfig}
\usepackage{epsf}
\usepackage[usenames]{color}

\def\half{\frac{1}{2}}

\def\Bn {{\bf n}}
\def\Bk {{\bf k}}
\def\BS {{\bf S}}

\def\re#1{(\ref{eq:#1})} 
   \def\lb{\lbrace}          \def\rb{\rbrace}
       
   \def\lan{\langle}         \def\ran{\rangle}
   \def\({\left(}            \def\){\right)}
   \def\[{\left[}            \def\]{\right]}

\begin{document}

\preprint{ITP-UU-07/67}

\title{Achieving the N\'eel state in an optical lattice}

\author{Arnaud Koetsier}
\email{koetsier@phys.uu.nl}

\affiliation{Institute for Theoretical Physics, Utrecht University, Leuvenlaan
4, 3584 CE Utrecht, The Netherlands}

\author{R.~A. Duine}

\affiliation{Institute for Theoretical Physics, Utrecht
University, Leuvenlaan 4, 3584 CE Utrecht, The Netherlands}

\author{Immanuel Bloch}

\affiliation{Institut f\"ur Physik, Johannes
Gutenberg-Universit\"at, 55099 Mainz, Germany}

\author{H.~T.~C. Stoof}

\affiliation{Institute for Theoretical Physics, Utrecht University, Leuvenlaan
4, 3584 CE Utrecht, The Netherlands}

\date{\today}

\begin{abstract}
We theoretically study the possibility of reaching the antiferromagnetic phase
of the Hubbard model by starting from a normal gas of trapped fermionic atoms
and adiabatically ramping up an optical lattice. Requirements on the initial
temperature and the number of atoms are determined for a three dimensional
square lattice by evaluating the N\'eel state entropy, taking into account
fluctuations around the mean-field solution. We find that these fluctuations
place important limitations on adiabatically reaching the N\'eel state.
\end{abstract}


\maketitle

\section{Introduction.}
An optical lattice is a regular periodic potential for neutral cold atoms
\cite{jessen1996} which enables the controlled experimental exploration of
paradigmatic ideas and models from condensed-matter physics. This is because
cold atomic gases generally allow for a great deal of experimental tunability.
For example, Feshbach scattering resonances allow for the interaction strength
to be experimentally varied over a considerable range
\cite{stwalley1976,tiesinga1993}. Other quantities that may be altered include
temperature, density, and strength and shape of the trapping potential. In
particular, an optical-lattice potential plays the role of the ion-lattice
potential encountered in electronic solid-state physics. The energy bands
resulting from this periodic potential lead to a quenching of the kinetic
energy of the atoms with respect to their interaction energy, enabling the
exploration of strongly-correlated phases that play a significant role in
condensed-matter physics.

An important model that can be studied experimentally with cold atoms is the
single-band Hubbard model, which consists of interacting fermions in the
tight-binding approximation. The Hubbard Hamiltonian is realized by cold atoms
in an optical lattice when the potential is strong enough so that only the
lowest-energy band is populated \cite{jaksch1998}. For bosonic atoms one then
commonly refers to this model as the Bose-Hubbard model. The theoretically
predicted Mott-insulator-to-superfluid phase transition \cite{fischer1989} for
this model has indeed been observed experimentally \cite{greiner2002}.

The fermionic Hubbard model, referred to simply as the Hubbard model, is
important in the context of high-temperature superconductivity
\cite{bednorz1986,Hofstetter2002} and has also been realized with cold atoms
\cite{kohl2004}. At half filling, corresponding to one particle per lattice
site, the ground state of this model is antiferromagnetic, i.e., a
N\'eel-ordered state, for strong enough on-site interactions. As the filling
factor is reduced by doping, the system is conjectured to undergo a quantum
phase transition to a $d$-wave superconducting state \cite{LeeNagaosaWen2006}.
A theoretical proof of the existence of $d$-wave superconductivity in the
Hubbard model is still lacking and would be a major step towards understanding
the superconducting state of the cuprates. With the recent experimental
advances in the field of ultracold atoms, an experimental exploration of this
issue is within reach.

In view of this motivation, a significant problem is determining how the N\'eel
state of the Hubbard model can be reached experimentally. In this paper, we
study theoretically the process of adiabatically turning on the optical lattice
\cite{GeorgesHassan2005,blakie2007}, with the goal of determining the
conditions required for an initially trapped balanced two-component Fermi gas
with repulsive interactions to reach the N\'eel state in the lattice.
Experimentally, the presence of antiferromagnetic order in this cold-atom
experiment can be subsequently detected from shot-noise correlations in the
density distribution \cite{altman2004,rom2006}.

\begin{figure}[b]
\begin{center}
\includegraphics[width=0.95\columnwidth]{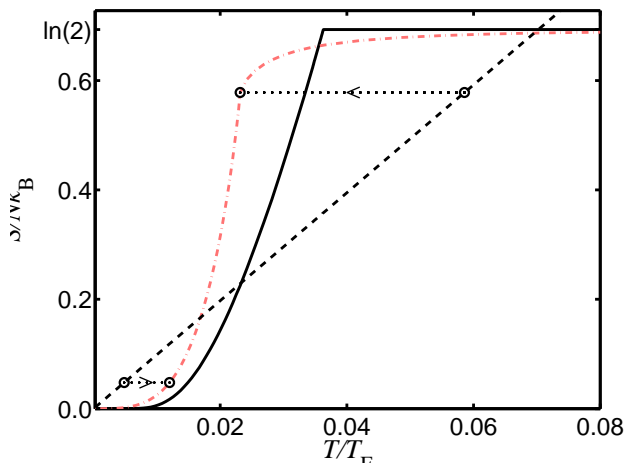}
 \bf \caption{ \rm (Color online). The entropy per particle in the harmonic trapping potential
only (dashed line), in a lattice of depth $V_0=6.5 E_{\rm R}$ ($E_{\rm R}$ is
the recoil energy) from single-site mean-field theory (solid curve) and with
fluctuations (dash-dotted curve), where $T_{\rm F}$ is the Fermi temperature in
the trap. The horizontal dotted lines illustrate cooling and heating into the
N\'{e}el state at constant entropy by starting in the harmonic trap and
adiabatically turning on the lattice.} \label{fig1}
\end{center}
\end{figure}

Our results are summarized in Fig.~\ref{fig1}. For initial temperatures lower
than $T_{\rm F}$, the Fermi temperature in the trap, the entropy per particle
in the trap depends linearly on temperature as is shown by the dashed line. The
optical lattice is then turned on adiabatically and to determine the final
temperature of the gas we need the entropy per atom in the lattice. For a
sufficiently smooth trapping potential such that the tunneling does not become
site-dependent, the only effect of the trap is to place a restriction on the
total number of particles which we discuss later and, other than this, we may
neglect the trap for calculations in the lattice. Since we consider balanced
gases here, we will at sufficiently low temperatures first enter the Mott phase
with one particle per site, and the subsequent evolution of the gas is then
described by the Heisenberg model for the spins alone. The result from the
usual mean-field theory is shown for a lattice depth of $6.5E_{\rm R}$ (where
$E_{\rm R}$ is the recoil energy) by the black curve, and is equal to $k_{\rm
B}\ln(2)$ everywhere above the critical temperature $T_c$. Since entropy is
conserved in adiabatic processes, the final temperature is simply the
temperature at which the final entropy in the lattice equals the initial
entropy in the trap. Two such processes are shown by the dotted lines for
different initial temperatures demonstrating that the gas is sometimes heated
and not cooled by the lattice. Nevertheless, mean-field theory leads to the
intuitive result that as long as the entropy per particle in the initial state
is less than $k_{\rm B} \ln(2)$, which is the maximum entropy of the Heisenberg
model, the N\'eel state is always reached by adiabatically turning on the
optical lattice.

The inclusion of fluctuations leads however to a more restrictive condition. To
probe the effect of fluctuations, we present an improved mean-field theory
which produces a temperature-dependent entropy above $T_c$, as seen from the
inset of Fig.~\ref{fig2}. Although this approach is exact at high temperatures,
it fails to account for spin waves present at low temperatures and for critical
phenomena near $T_c$. By further extending the improved mean-field theory to
reproduce the correct critical and low temperature behavior due to
fluctuations, we are able to determine the entropy in the lattice for all
temperatures (red curve in Fig.~\ref{fig1}). In particular, we find that
fluctuations lower the entropy of the atoms in the square lattice at $T_c$ as
\begin{equation}
  \label{eq:criticalS}
  S(T=T_c) \simeq Nk_{\rm B}\ln(2) - \frac{3N J^2}{32 k_{\rm B} T_c^2 (3\nu - 1)}~,
\end{equation}
where $\nu$ is the critical exponent of the correlation length $\xi$. For the
case of three dimensions, $\nu=0.63$ \cite{zinnjustinbook}. As a result, the
initial temperature required to reach the N\'eel state is more than 20\% lower
than that found from the usual mean-field theory, but fortunately remains
experimentally accessible. For example, with ${}^{40}$K atoms and a final
lattice depth of $8E_{\rm R}$ the N\'eel state is achieved when the final
temperature in the lattice is $0.012T_{\rm F}$, which can be obtained with an
initial temperature of $0.059T_{\rm F}$.

\section{Single-site mean-field theory.}
The Hamiltonian for the Hubbard model is given by
\begin{equation}
\label{eq:hubbardham}
  H = - t \sum_{\sigma} \sum_{\langle jj'\rangle}
               c_{j,\sigma}^\dagger c_{j',\sigma}
  + U \sum_j c_{j,\uparrow}^\dagger c_{j,\downarrow}^\dagger
             c_{j,\downarrow} c_{j,\uparrow}~,
\end{equation}
in terms of fermionic creation and annihilation operators, denoted by
$c^\dagger_{j,\sigma}$ and $c_{j,\sigma}$, respectively, where $\sigma$ labels
the two hyperfine spin states $|\unskip\mkern-7.5mu\uparrow \rangle$ or
$|\unskip\mkern-7.5mu\downarrow \rangle$ of the atoms. In the first term of
this expression, the sum over lattice sites labeled by indices $j$ and $j'$ is
over nearest neighbors only and proportional to the hopping amplitude given by
\begin{equation}
  t = \frac{4 E^{\mathrm{R}}}{\sqrt{\pi}}
    \(\frac{V_0}{E^{\mathrm{R}}}\)^\frac{3}{4}
    e^{-2\sqrt{V_0 / E^{\mathrm{R}}}}.
\end{equation}
Here, $V_0>0$ is the depth of the optical lattice potential defined by
\begin{equation}
  V({\bf x})=V_0[\cos^2(2\pi x/\lambda) + \cos^2(2\pi y/\lambda) + \cos^2(2\pi z/\lambda)],
\end{equation}
where $\lambda$ is the wavelength of the lattice lasers. The second term in the
Hamiltonian corresponds to an on-site interaction of the strength given in the
harmonic approximation by
\begin{equation}
  U=4\pi a\sqrt{\frac{\hbar}{\lambda^3}}\left(\frac{8 V_0^3}{m}\right)^\frac{1}{4},
\end{equation}
where $a$ is the $s$-wave scattering length which is equal to $174a_0$ for
${}^{40}K$. It is well-known \cite{auerbachbook} that at half filling and in
the limit that $U \gg t$ the ground state of the Hubbard model is
antiferromagnetic and that, for $k_{\rm B}T \ll U$, its low-lying excitations
are described by the effective Heisenberg Hamiltonian
\begin{equation}
\label{eq:spinham}
  H = \frac{J}{2} \sum_{\langle jk\rangle}
                     \BS_j \cdot \BS_k~,
\end{equation}
with $\BS$ being one half times the vector of Pauli matrices. The exchange
constant $J=4t^2/U$ arises from the superexchange mechanism. That is, the
system can lower its energy by virtual nearest-neighbor hops only when there is
antiferromagnetic ordering.

Within the usual mean-field analysis of the effective Hamiltonian in
Eq.~\re{spinham}, the total entropy for $N$ atoms in the optical lattice is
given by
\begin{equation}
\label{eq:entropymf}
  S = -\frac{\partial F_{\rm L}(\lan\Bn\ran)}{\partial T}~,
\end{equation}
where $F_{\rm L}$ is the Landau free energy,
\begin{equation}
\label{eq:freeenergymftheory}
  F_{\rm L} ({\bf n}) =N\left\{ \frac{z J |{\bf n}|^2}{2}
                       - k_{\rm B} T \ln \left[2 \cosh \left(
  \frac{z J|{\bf n}|}{ k_{\rm B} T}\right)\right] \right\}~,
\end{equation}
in terms of the staggered, or N\'eel, order parameter ${\bf n} = (-)^j \langle
\BS_j \rangle$ for the phase transition to the antiferromagnetic state. In the
expression for the free energy, $z=6$ is the number of nearest neighbors for a
three-dimensional simple square lattice on which we focus here, $k_{\rm B} T$
is the thermal energy, and $\langle {\bf n} \rangle$ is the equilibrium value
of the order parameter determined from
\begin{equation}
\label{eq:expecvalueopmftheory}
  \left.\frac{\partial F_{\rm L} ({\bf n})}{\partial {\bf
  n}}\right|_{\Bn=\lan\Bn\ran}=0~.
\end{equation}
It is nonzero below a critical temperature $k_{\rm B} T_c = J z/4 = (3/2)J$.
After solving Eq.~\re{expecvalueopmftheory} the entropy is determined using
Eq.~\re{entropymf}. The results for $S$ and $\lan\Bn\ran$ obtained in this way
are plotted as solid black curves in Figs.~\ref{fig1} and \ref{fig2}.

The entropy $S_{\rm FG}$ of the initial normal state before ramping up the
optical lattice is the entropy of a trapped ideal Fermi gas. It is most
conveniently determined from the grand potential
\begin{equation}
\label{eq:grandpotidealfermigas}
  \Omega (\mu,T) = - k_{\rm B } T \int_0^\infty d\epsilon \rho
  (\epsilon) \ln \left[ 1+ e^{-(\epsilon - \mu)/k_{\rm B} T}\right]~,
\end{equation}
where $\mu$ is the chemical potential, and the effect of the harmonic trapping
potential with the effectively isotropic frequency
$\omega=(\omega_x\omega_y\omega_z)^{1/3}$ is incorporated via the density of
states $\rho (\epsilon) = \epsilon^2/(\hbar \omega)^3$ of the atoms. The
entropy at fixed total particle number $N(\mu) = -\partial \Omega/\partial \mu$
is then given by $ S_{\rm FG} = -\partial \Omega/\partial T|_{\mu (N)}$. At
temperatures much lower than the Fermi temperature in the trap, given by
$T_{\rm F} = (3N)^{1/3}\hbar\omega/k_{\rm B}$, we find in this manner that
\cite{Carr04} $S_{\rm FG} = Nk_{\rm B}  \pi^2 T/T_{\rm F}$.
Now, by equating the final and initial entropies we calculate the temperature
of the Heisenberg spin system that results after adiabatically turning on the
optical lattice, in terms of the initial temperature of the trapped Fermi gas.

From the expression for the free energy, Eq.~\re{freeenergymftheory} we
immediately see that $S = Nk_{\rm B}\ln(2)$ for all temperatures $T>T_c$, as
was shown in Fig.~\ref{fig1}. Although this is the correct high-temperature
limit of the entropy, temperature dependence will lower the entropy at $T_c$
and therefore lower the initial temperature required to achieve the N\'eel
state. To obtain the temperature dependence above $T_c$, we must thus go beyond
single-site mean-field theory to include fluctuations. The simplest such model
described below incorporates the interaction of a given site with one of its
neighbors exactly and treats interactions with the rest of the neighbors within
mean-field theory.

\section{Two-site mean-field theory.}
\begin{figure}[h]
\begin{center}
\includegraphics[width=0.95\columnwidth,bb=21 200 542 588]{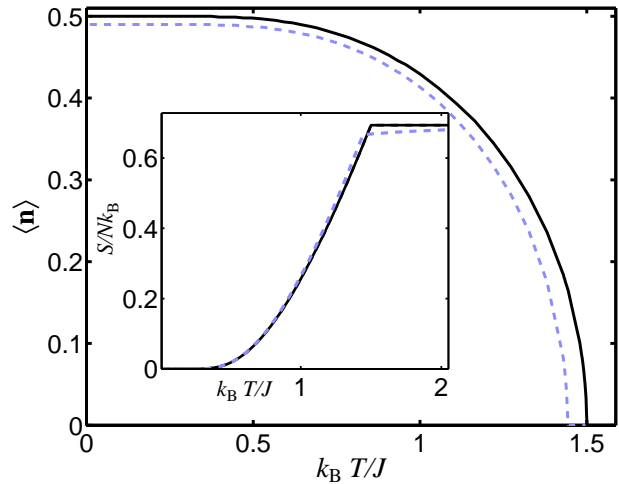}
\bf \caption{ \rm (Color online). The staggered magnetization $\lan\Bn\ran$ of
the single (solid curves) and two-site (dashed curves) mean-field theories, the
latter of which shows depletion at $T=0$ and a lowering in $T_c$. The entropy
of both theories is plotted in the inset. Above $T_c$ we see that the entropy
of the two-site theory is temperature dependent.} \label{fig2}
\end{center}
\end{figure}

The two-site Hamiltonian for neighboring sites labeled ``1'' and ``2'' is given
by
\begin{equation} \label{eq:2siteham}
  H =   J\BS_1\cdot\BS_2
       + J(z-1)|\Bn| (\BS_1^z - \BS_2^z)
       + J(z-1)|\Bn|^2~,
\end{equation}
where the last term is a correction to avoid double counting of mean-field
effects. Diagonalizing this Hamiltonian we obtain the free energy
\begin{equation}
\label{eq:freeenergy2sitetheory}
     \begin{aligned}[b]
 F_{\rm L} ({\bf n}) =&
     N\biggl\lb\half(z-1)J|\Bn|^2 - \frac{1}{2\beta}\ln \biggl[2 e^{-\beta J/4}
     \\&
     +2 e^{\beta J/4} \cosh \( \frac{\beta J}{2} \sqrt{1 + 4(z-1)^2|{\bf n}|^2}\)\biggr]\biggr\rb~,
     \end{aligned}
\end{equation}
and find the entropy from Eq.~\re{entropymf} with the condition
Eq.~\re{expecvalueopmftheory} as in the single-site model. The results are
plotted in Fig.~\ref{fig2}, where we see that fluctuations lower the critical
temperature and also bring about a 2\% depletion of the order parameter which
is now less than 0.5 near $T=0$.

The two-site result carries the exact $1/T^2$ dependence of the entropy of the
Heisenberg model at high temperatures. Near $T=0$, however, the entropy is
still exponentially suppressed reflecting the energy cost of flipping a spin.
This exponential suppression is an artefact of the mean-field approximation
that ignores the Goldstone modes which are present in the symmetry-broken
phase. Furthermore, critical behavior cannot be properly accounted for by a
one-, two- or higher-site model since, near the onset of N\'eel order, critical
fluctuations extend throughout the entire lattice so one would in principle
need to include all sites exactly. To overcome these shortcomings, we extend
our two-site model below to all temperatures.

\section{Fluctuations.}
The two-site mean-field theory produces the correct normal-state entropy
behavior in the high-temperature limit,
\begin{equation}
  S(T\gg T_c) = Nk_{\rm B} \[\ln(2) -\frac{3J^2}{64k_{\rm B}^2T^2}\]~.
\end{equation}
In the low-temperature regime, the entropy is determined from spin-wave
fluctuations prevalent near $T=0$ which give a black-body-like entropy,
\begin{equation}
\label{eq:magnonentropy}
  S(T\ll T_c)=Nk_{\rm B} \frac{4\pi^2}{45}\(\frac{k_{\rm B}T}{2\sqrt{3}J\lan\Bn\ran}\)^3~.
\end{equation}

The continuous interpolation between these two regimes has the additional
constraint that, near $T_c$, we should obtain the correct critical behavior of
the antiferromagnet, namely, the correct universal ratio of the amplitudes
above and below the phase transition $A^+/A^-$ and correct critical exponent
$d\nu - 1$ where
\begin{equation}
  S(T\simeq T_c) = S(T_c)\pm A^\pm|t|^{d\nu - 1},\quad t=(T-T_c)/T_c\rightarrow 0^\pm~.
\end{equation}
This follows from the fact that the singular part of the free energy density
behaves as $F^{\pm}/\xi^d$, where the correlation length diverges like
$\xi\sim|t|^{-\nu}$ as $t\rightarrow 0$. Explicit expressions for the entropy
embodying the correct behavior in the low-, high- and critical temperature
regimes are presented in the Appendix of this paper, and plotted as the red
curve in Fig.~\ref{fig1} for $d=3$ using $A^+/A^-\simeq0.54$ and $\nu=0.63$
\cite{zinnjustinbook}, and the N\'eel temperature of $T_c = 0.957 k_{\rm B}/J$
\cite{Staudt2000}. Their value at $T_c$ leads to the central result of this
paper, namely, Eq.~(\ref{eq:criticalS}) which specifies the initial entropy
required to reach the N\'eel state.

\section{Discussion and Conclusions.}
As briefly mentioned earlier, there is a limit on the total number of atoms in
the trap, beyond which at low temperatures it is energetically more favorable
to doubly occupy sites in the center of the trap, thereby destroying the
antiferromagnetic state, rather than singly occupying outlying sites where the
trap potential is larger than $U$. Thus, insisting that the system end up in
the Mott-insulator state with one particle per site entails the upper bound, $N
\leq N_{\rm max}=(4\pi/3)(8U/m\omega^2\lambda^2)^{3/2}$, where $m$ is the mass
of the atoms and $\lambda$ is the wavelength of the lattice lasers. For
${}^{40}$K atoms in a lattice with a wavelength $\lambda=755$ nm and depth
$8E_{\rm R}$, and with a harmonic trap frequency $\omega/2\pi = 50$ Hz, $N_{\rm
max} \simeq 2\times 10^6$ which is well above the typical number of atoms in
experiments.

We have also attempted to determine the effect that fluctuations have on the
entropy in a more microscopic manner by studying gaussian fluctuations about
the mean-field $\lan\Bn\ran$ for the single-site mean-field theory in the
low-temperature regime. But such a random-phase approximation has severe
complications related to the fact that $\lan\Bn\ran$ enters in the magnon
dispersion as $\omega^{\rm M}_{\Bk}\propto \lan|\Bn|\ran|\Bk|$. Hence, as can
be already seen from Eq.~\re{magnonentropy}, the contribution of the magnons to
the entropy diverges when $\lan\Bn\ran\rightarrow 0$ near $T_c$. One way to
potentially resolve this issue is to start from the Hubbard Hamiltonian
Eq.~\re{hubbardham} directly but such an analysis is involved \cite{dupuis04}
and has yet to be carried out.

In the above, we have focused on the $d=3$ case. Whilst our results can easily
be extended to the $d=2$ case, a more pertinent way to reach the
two-dimensional antiferromagnet most relevant to high-temperature
superconductors, would be to adiabatically prepare a three-dimensional N\'eel
state, as explained in this paper, and then decrease the tunneling in one
direction by changing the intensity of one of the lattice lasers. In this way,
the three-dimensional system is changed into a stack of pancakes of atoms in
the two-dimensional N\'eel state. Furthermore, studying doped optical lattices
made by introducing a small imbalance in the initial state may shed some light
on the physics of high-temperature superconductors and would be an exciting
direction for future research.

\begin{acknowledgments}
We would like to thank Gerard Barkema, Dries van Oosten and Randy Hulet for
helpful discussions. This work is supported by the Stichting voor Fundamenteel
Onderzoek der Materie (FOM), the Nederlandse Organisatie voor Wetenschaplijk
Onderzoek (NWO), and the Deutsche Forschungsgemeinschaft (DFG).
\end{acknowledgments}

\appendix*
\section{Entropy Formulas}
For temperatures above $T_c$ we use
    \begin{equation*}
      \frac{S(T\geq T_c)}{Nk_{\rm B}} \simeq
          \alpha_1 \[\(\frac{T-T_c}{T}\)^{\kappa}-1+ \frac{\kappa T_c}{T}\]+\ln(2)~,
    \end{equation*}
with $\alpha_1 = 3J^2/32\kappa(\kappa - 1)k_{\rm B}^2 T_c^2$ and $\kappa = 3\nu
- 1\simeq 0.89$ \cite{zinnjustinbook}. The first term embodies the correct
critical behavior whereas the remaining terms are present to recover the
correct high-temperature limit. Below $T_c$, however, we have
    \begin{equation*}
    \begin{aligned}[b]
      \frac{S(T\leq T_c)}{Nk_{\rm B}} =
      -\alpha_2 &\bigg[\(\frac{T_c - T}{T_c}\)^\kappa - 1 + \kappa \frac{T}{T_c}
      \\&- \frac{\kappa(\kappa - 1)}{2} \frac{T^2}{T_c^2}\bigg]
      + \beta_0 \frac{T^3}{T_c^3}
      + \beta_1 \frac{T^4}{T_c^4}~,
      \end{aligned}
    \end{equation*}
where
    \begin{equation*}
    \begin{aligned}[t]
     \alpha_2 =& \begin{aligned}[t]
         \frac{6}{(\kappa-1)(\kappa-2)(\kappa-3)}&
         \bigg(\frac{4\pi^2 k_{\rm B}^3 T_c^3}{135\sqrt{3}J^3} \\ & - \alpha_1(\kappa-1) + \beta_1  - \ln(2)\bigg)~;
         \end{aligned}\\
     \beta_0 =& \frac{\kappa}{(\kappa-3)}
         \bigg(\frac{4\pi^2 k_{\rm B}^3 T_c^3}{45\sqrt{3}\kappa J^3} + \alpha_1 (\kappa-1) - \beta_1  + \ln(2)\bigg)~;
         \\
     \beta_1 =& \ln 2 - J^2\frac{6(A^+/A^- + 1) + \kappa(\kappa-5)}{64\kappa k_{\rm B}^2 T_c^2 A^+/A^-}
      - \frac{4\pi^2 k_{\rm B}^3 T_c^3}{135\sqrt{3}J^3}~.
    \end{aligned}
    \end{equation*}
The first and last terms in $S(T\leq T_c)$ embody the critical phenomena and
allow for the continuous interpolation with $S(T\geq T_c)$ respectively,
whereas the remaining terms are included to retrieve the correct
low-temperature behavior.


\end{document}